\documentclass[conference]{IEEEtran}
\IEEEoverridecommandlockouts
\usepackage{cite}
\usepackage{amsmath,amssymb,amsfonts}
\usepackage{amsthm}
\usepackage{algorithm}
\usepackage{algorithmic}
\usepackage{graphicx}
\usepackage{caption}
\usepackage{textcomp}
\usepackage{pgfgantt}
\usepackage{listings}
\usepackage{rotating} 
\usepackage{sidecap}
\usepackage{xcolor}
\usepackage{enumitem}
\usepackage{cite}
\usepackage{comment}
\usepackage{stackrel,amssymb,amsmath,turnstile,mathtools}
\usepackage{url}
\usepackage{hyperref}
\usepackage{tcolorbox}
\usepackage{multirow}
\usepackage[utf8]{inputenc}
\usepackage{amsmath}
\usepackage{lineno}
\usepackage{amsfonts}
\usepackage{ mathrsfs }
\usepackage{anyfontsize}

\def\BibTeX{{\rm B\kern-.05em{\sc i\kern-.025em b}\kern-.08em
    T\kern-.1667em\lower.7ex\hbox{E}\kern-.125emX}}

\lstdefinestyle{codestyle}{
  commentstyle=\color{codegreen},
  keywordstyle=\color{magenta},
  numberstyle=\tiny\color{black},
  stringstyle=\color{purple},
  basicstyle=\ttfamily\footnotesize,
  breakatwhitespace=false,         
  breaklines=true,                 
  captionpos=b,                    
  keepspaces=true,                 
  numbers=left,                    
  numbersep=5pt,                  
  showspaces=false,                
  showstringspaces=false,
  showtabs=false,
  emph={assert,uint64_t,uint32_t,assume},
  emphstyle=\color{magenta},
  frame = single,
  framexleftmargin=8pt,
  tabsize=2,
  xleftmargin= 18pt,
  xrightmargin=18pt
}
\newlist{RQ}{enumerate}{1}
\setlist[RQ, 1]{label = \textbf{RQ \arabic*},leftmargin=20pt,rightmargin=10pt}

\DeclareSymbolFont{symbolsC}{U}{pxsyc}{m}{n}
\DeclareMathSymbol{\circledless}{\mathrel}{symbolsC}{82}
\newcommand{\cou}[1]{\texttt{#1}}
\newcommand{\expression}{\textit{\texttt{E\,}}}
\newcommand{\bexpression}{\textit{\texttt{B\,}}}
\newcommand{\logic}{\textit{\texttt{L\,}}}

\definecolor{myyellow}{RGB}{245,205,0}

\let\emptyset\varnothing

\renewcommand{\baselinestretch}{0.966}

\begin{document}
\clearpage
\pagestyle{plain}
\pagenumbering{arabic}

\title{Incremental Symbolic Bounded Model Checking of Software Using Interval Methods via Contractors}%\thanks{The work in this paper is partially funded by the EPSRC grants  EP/T026995/1, EP/V000497/1, and EU H2020 ELEGANT 957286.  The first author acknowledges the scholarship received from the King Saud University, Saudi Arabia.}}

\author{\IEEEauthorblockN{Mohannad Aldughaim, Kaled Alshmrany, Rafael Menezes, Alexandru Stancu, and Lucas C. Cordeiro}
\IEEEauthorblockA{\textit{University of Manchester, Manchester, UK}}} 

%email address or ORCID
%}
%\and
%\IEEEauthorblockN{Kaled Alshmrany}
%\and
%\IEEEauthorblockN{Rafael Menezes}
%\and
%\IEEEauthorblockN{Alexandru Stancu}
%\and
%\IEEEauthorblockN{Konstantin Korovin}
%\and
%\IEEEauthorblockN{Lucas C. Cordeiro}
%Manchester, UK \\

\maketitle

\begin{abstract}
Bounded model checking (BMC) is vital for finding program property violations. For unsafe programs, BMC can quickly find an execution path from an initial state to the violated state that refutes a given safety property. However, BMC techniques struggle to falsify programs that contain loops. BMC needs to incrementally unfold the program loops up to the bound $k$, exposing the property violation, which can thus lead to exploring a considerable state space. Here, we describe and evaluate the first verification method based on interval methods via contractors to reduce the domains of variables representing the search space. This reduction is based on the specified property modeled as functions representing the contractor constraints. In particular, we exploit interval methods via contractors to incrementally analyze the program loop variables and contract the domain where the property is guaranteed to hold to prune the search exploration, thus reducing resource consumption aggressively. Experimental results demonstrate the efficiency and efficacy of our proposed approach over a large set of benchmarks, including $7044$ verification tasks, compared with state-of-the-art BMC tools. Our proposed method can reduce memory usage up to $75$\% while verifying $1$\% more verification tasks.
\end{abstract}


%-----------------------------------------------
\section{Introduction}
%-----------------------------------------------
\theoremstyle{definition}
\newtheorem{definition}{Definition}

Software model checking prevents software errors that could cause financial losses or risk individuals' well-being \cite{Clarke2012ModelProblem}. Developed originally by Clarke and Emerson~\cite{Clarke2008} and by Queille and Sifakis \cite{10.1007/3-540-11494-7_22}, Model Checking models the software into states that will be explored exhaustively; it is considered one of the most successful methods for its use in research and industry \cite{Clarke2012ModelProblem}. Model-checking started with mainly hardware verification in mind, and over the years, researchers developed the technique to include many aspects of software verification. However, one of the critical problems is the state-space explosion problem~\cite{Clarke2012ModelProblem}. The growth rate of states is exponential in the number of variables and processes. With fairly complex software systems, the number of states it will produce can grow out of proportion, which, in turn, will take a substantial amount of memory and time to perform the verification process automatically. 
 
There exist studies that tackled the state explosion problem using various efficient data structures to represent the system's states~\cite{McMillan1993,1676819,10.1007/3-540-56922-7_34,10.1007/10722167_15,10.1007/3-540-49059-0_14}. 
One of the first techniques is Symbolic Model Checking with Binary Decision Diagrams (BDDs)~\cite{McMillan1993}\cite{1676819}, which is performed using a fixed-point algorithm. Here, rather than listing each state individually, the states are represented using a BDD, exponentially smaller than explicitly enumerating the entire state-space and performing explicit-state model checking. Peled~\cite{10.1007/3-540-56922-7_34} introduced Partial Order Reduction (POR) techniques to reduce state-space exploration. This technique is best suited for concurrent systems as it takes advantage of the commutative transitions that result in the same state. Another successful attempt was Counterexample-Guided Abstraction Refinement (CEGAR) made by Clarke et al. \cite{10.1007/10722167_15}. This method converts the code to be verified into an abstract model. Then it refines this model according to the property being verified. This method keeps the state space small using abstraction functions that determine each variable's level of abstraction. 

Biere et al. in $1999$ \cite{10.1007/3-540-49059-0_14} introduced Bounded Model Checking (BMC), which uses Boolean Satisfiability (SAT) or Satisfiability Modulo Theories (SMT) solvers' speed to find a counterexample of a fixed length, reducing the state-space size exploration. BMC aims to find errors rather than prove the absence of them~\cite{Biere2003}. Some approaches aim to under-approximate and widen the system's model based on the proof of unsatisfiability extracted from SMT solvers~\cite{DBLP:phd/ethos/Cordeiro11}. And some approaches over-approximate and then narrow the search, such as property-directed reachability (PDR)~\cite{cimatti2012software}. PDR is a derivative of the IC3 algorithm that has been used successfully to verify hardware~\cite{hoder2012generalized}.  

Here, we advance the state-of-the-art in software model checking and introduce a novel verification method to reduce the search space exploration using interval methods via contractors~\cite{Jaulin2001AppliedIA}. Abstract interpretation techniques exist to compute intervals~\cite{cousot1977abstract}. If we compare interval analysis and methods with abstract interpretation techniques, they share the same formal definitions and operations over intervals. However, they differ regarding interval methods since we can use \textit{forward} and \textit{backward} operations to ``contract'' the search space exploration from the program's entry point to the property being verified and vice-versa. Therefore, contractors make interval methods more practical by reducing the domain of the program variables using off-the-shelf solvers targeted at constraint satisfaction problems (CSP)~\cite{ibex}. Interval methods applications in robotics, medicine, and finance, showed that it could prove the existence of a solution while taking into account all sources of uncertainty~\cite{MUSTAFA2018160}.

Intuitively, our approach works as follows: first, we model the software properties as constraints. Second, we analyze the intervals for each variable. Third, we apply contractors to analyze those variables' search space, identify where a property is guaranteed to hold, and remove the states associated with identified values. What remains is a smaller upper bound region where constraints are satisfied. Finally, we provide those regions with the BMC procedure to perform the automated symbolic verification. Experimental results show how our approach improved verification tasks' scores while lowering resource consumption. This work significantly extends our technical report~\cite{abs-2012-11245}, which now considers the full implementation and extensive experimental evaluation of our proposed method.

%----------------------------------------------------------------
\noindent \textbf{Contributions.} 
%----------------------------------------------------------------
The main original contributions of this paper can be summarised as follows:

\begin{itemize}
\item We develop and evaluate a novel verification method based on interval methods via contractors, which contracts the domains of variables representing the search space. This contraction is based on the specified property modelled as functions representing the contractor constraints.
\item We prune the search space not to include the verified property region, which will always hold, thus reducing time and memory usage. We perform this step using the contractor to reduce the guaranteed area inside the solution.
\item Experimental results demonstrate the efficiency and efficacy of our proposed approach over a large set of benchmarks compared with state-of-the-art BMC tools, where we can reduce memory usage up to $75$\% while verifying $1$\% more verification tasks.
\end{itemize} 

%----------------------------------------------------------------
\section{Preliminaries}
\label{sec:Background}
%----------------------------------------------------------------

%\textcolor{cyan}{we need to ensure that we refer to each definition throughout the text.}

%-----------------------------------------------------
\subsection{Soundness and Completeness}
%-----------------------------------------------------

%
\begin{definition}[\textbf{\emph{Soundness}}]
\label{def:soundness}
For a verifier to be considered \emph{sound}, it must reason about all program executions. Thus, if the verifier result concludes that there are no bugs, i.e., the verification outcome is TRUE, then the program is \textit{safe} concerning the specified property. Formally, we can define \emph{soundness} with the following equation: 
\begin{equation}
\label{soundness}
    \forall\; \texttt{p} \in \mathbb{L},\; \texttt{verifier(p)} = \textbf{TRUE} \implies \texttt{p} \;\text{satisfies}\; \mathscr{P}
\end{equation} 
\noindent where $\mathbb{L}$ is the set of all programs, the program \texttt{verifier} is sound with respect to property $\mathscr{P}$~\cite{rival2020introduction}
\end{definition}

\begin{definition}[\textbf{\emph{Completeness}}]
\label{def:completeness}
For a verifier to be considered \emph{complete}, it must have a concrete program execution that results in a property violation. Thus, if the verifier reports a bug, then the program is \textit{unsafe} concerning the specified property. Similar to soundness~\ref{soundness}, we can define completeness with the following equation:
\begin{equation}
    \forall\; \texttt{p} \in \mathbb{L},\; \texttt{verifier(p)} = \textbf{TRUE} \impliedby \texttt{p} \;\text{satisfies}\; \mathscr{P}
\end{equation}
\end{definition}

%----------------------------------------------------------------
\subsection{Incremental Bounded Model Checking (BMC) of Software}
\label{sec:BMC}
%----------------------------------------------------------------

In BMC, the program is modeled as a state transition system (TS), which is derived from its control-flow graph~\cite{Muchnick1997AdvancedImplementation}, and then converted to a \emph{Static Single Alignment} form (SSA). Each control graph node will be converted to either an assignment or a conditional statement converted to a guard. Each edge represents a change in the program's control location \cite{Cordeiro2012SMT-basedSoftware}.
When modeling the program, a Kripke structure \cite{Kripke1963SemanticalCalculi} is used as TS $M=(S,T,S_0)$, which is an abstract machine that has a set of states $S$, initial states $S_0 \subseteq S$, and transition relation $T\subseteq S\times S$. The set of states $S=\{s_0,s_1,...s_n\}: n\in \mathbb{N}$ contains all the states. Each state has the value of each variable in the program and a program counter $pc$. Each transition is denoted by $\gamma(s_i,s_{i+1})\in T$, where it represents a logical formula that encodes all the changes in variables and $pc$ from $s_i$ to $s_{i+1}$.
Next is to compute a \emph{Verification Condition (VC)} denoted by $\Psi$, a quantifier-free formula in a decidable subset of first-order logic. Given a transition system $M$, a property $\phi$, and a bound $k$, BMC unrolls the system $k$ times to produce $\Psi$ such that $\Psi$ is satisfiable \emph{iff} $\phi$ has a counterexample of length $k$ or less.

\begin{equation}
\Psi_k = I(s_0)\wedge \bigvee\limits_{i=0}^{k}{\bigwedge\limits_{j=0}^{i-1}\gamma(s_j,s_{j+1})\wedge\neg\phi(s_i) }
\label{eqBMC}
\end{equation}

In the logical formula \eqref{eqBMC}, $I$ denotes the set of initial states of $M$, $\gamma(s_j,s_{j+1})$ denotes the relation between two states in $M$. $\phi$ denotes the safety properties that should not be violated. Hence, $I(s_0)\wedge \bigwedge_{j=0}^{i-1}\gamma(s_j,s_{j+1})$ represents the execution of $M$ of length $i$ and if there exists some $i \leq k$ such that it satisfies $\Psi _k$ at time-step $i$ there exists a state in which $\phi$ is violated. Then $\Psi _k$ is given to an SMT solver to be checked for satisfiability. If it is satisfiable, then the SMT solver will provide an assignment that satisfies $\Psi _k$. With this assignment, the counterexample is constructed using the values extracted from the program variables. For a property $\phi$, a counterexample consists of a sequence of states $\{s_0,s_1,..,s_k\} | s_0 \in S_0$ and $s_i \in S|0\leq i< k$ and $\gamma(s_i,s_{i+1})$. If $\Psi _k$ is unsatisfiable, no error state is reachable in $k$ steps or less; therefore, no property was violated. 
Tools that implement BMC for software~\cite{Cordeiro2012SMT-basedSoftware} produce two quantifier-free formulae, represented by $C$ and $P$, which encode the constraints and properties, resp. Formula $C$ serves as the first part of $\Psi _k$, which is $I(s_0)\wedge \bigvee_{i=0}^{k} \bigwedge_{j=0}^{i-1}\gamma(s_j,s_{j+1})$. As for $\neg P$, it serves as the second part of $\Psi _k$ which is $\bigvee_{i=0}^{k}\neg\phi(s_i)$. Then the SMT solver checks $C\models_{{\mathcal{T}}} P$ in the form of $C\land \neg P$. 

BMC can also be used for soundness (cf. Definition~\ref{def:soundness}) under certain conditions. A completion threshold ($CT$) is used to provide soundness~\cite{Biere2003}. This threshold checks whether $k$ reaches all reachable states of $M$. This means a path of size $k$ that can reach the program's last state starting from the initial state. $CT_k$ can be defined as:
\begin{equation}
CT_k = I(s_0)\wedge \bigwedge\limits_{i=0}^{k} \gamma(s_i,s_{i+1}) \wedge \neg(s_k = s_n)
\end{equation}

\noindent where $n$ is the last reachable state of $M$. In BMC, if $\Psi_k \vee CT_k$ is UNSAT for a given $k$, then the program can be considered \textit{safe}. Finding the optimal value for $k$ is not always possible due to unbounded loops ($i$) or limited system resources ($ii$). Considering these limitations, two approaches can be used: \textit{k}-induction (for $i$) and incremental (for $ii$).

\begin{definition}[\textbf{\emph{Incremental-BMC}}]
\label{def:incrBMC} In this approach, the value of $k$ is incremented interactively until all states are reached (or resources are exhausted)~\cite{esbmc2018}. We can define the incremental-BMC for a program $P$ and bound $k$ as $Inc(P,k)$:
\begin{equation}
Inc(P,k) = \begin{cases}\text{P is unsafe}, &\Psi_k \text{ is SAT} \\
\text{P is correct}, &\Psi_k \vee CT_k \text{ is UNSAT} \\
Inc(P,k+1), &\mathrm{otherwise}.
\end{cases}
\end{equation}
\end{definition}

% \begin{comment}
% \begin{definition}[\textbf{\emph{\textit{k}-Induction}}]
% \label{kindBMC} This approach requires three steps: base case, forward condition and inductive step. Base case and forward condition are represented through $\Psi_k$ and $CT_k$, respectively. The inductive step (referred to as $S$) checks whether a property that holds for $k$ will imply that it also holds for any next $k$. To achieve this, $S$ is applied to a $M'$, which adds lambda transitions between every state \cite{esbmc2018} as follows:
% %
% \begin{equation}
%  I_k = \exists n. \bigwedge^{n + k - 1}_{i=n} (\phi(s_i)
% \wedge \gamma^{'}(s_i, s_{i+1})) \wedge \neg \phi(s_{n+k})
% \end{equation}

% \begin{equation}
% K(P,k) =  \begin{cases}\text{P has a bug}, &\Psi_k \text{ is SAT} \\
% \text{P is correct}, &\Psi_k \vee (CT_k \vee S_k) \text{ is UNSAT} \\
% K(P,k+1), &\mathrm{otherwise}.
% \end{cases}
% \end{equation}
% \end{definition}
% \end{comment}

%----------------------------------------------------------------
\subsection{Interval Analysis and Methods}
\label{sec:IAM}
%----------------------------------------------------------------

\begin{definition}[\textbf{\emph{Interval}}]
\label{Interval}
An \emph{interval} is a connected, closed subset of $\mathbb{R}$ denoted by $[x]$. It has a lower and an upper limits that are scalars denoted by $\underline{x}$ and $\overline{x}$, respectively, where $\underline{x},\overline{x} \in \mathbb{R}$.  An interval is defined as $[x]=[\underline{x},\overline{x}]\in \mathbb{IR}$, where $\mathbb{IR}$ is the set of all intervals of $\mathbb{R}$ \cite{Sainz2014ModalAnalysis}.
\end{definition}

 Here, we assume that $\underline{x}\leq\overline{x}$, nonetheless, in the field of \emph{Model Interval Analysis}~\cite{Sainz2014ModalAnalysis} it is possible to have $\underline{x}>\overline{x}$. However, this case is outside this paper's scope, which does not mean that the opposite is invalid as it is used in the field of \emph{Model Interval Analysis}~\cite{Sainz2014ModalAnalysis}. 
Each interval has a \emph{width} and \emph{center}, which are respectively defined as: $w([x])=\overline{x}-\underline{x}$ and $ c([x]) = \frac{\overline{x}+\underline{x}}{2}$, where \emph{width} and \emph{center} are $w$ and $c$, respectively. 
An interval is considered \emph{degenerate} or \emph{punctual} if its width is $0$, i.e., $w=0$.
Intervals can be extended to multiple dimensions. An interval in $n$ dimensions is a vector of intervals for each dimension. Such as $[\mathbf{x}] = [x_1]\times[x_2]\times...\times[x_n]$, where $[\mathbf{x}]$ is called a \emph{box} and $[\mathbf{x}]\in \mathbb{IR}^n$. A box is an axis-aligned hyperrectangle in $\mathbb{R}^n$.
\begin{comment}
Interval width and center are also extended to boxes as follows:
%
\begin{equation}
 w([\mathrm{x}]) = \max_{i \in \{1,...,n\}} w([x_i]),
    \label{eq:box width}
\end{equation}
\begin{equation}
c([\mathrm{x}]) = [c([x_1])\quad c([x_2])\quad ...\quad c([x_n])]^T.
    \label{eq:box center}
\end{equation}
The \emph{center} of a box is a vector of each interval center;  \emph{width} remains a scalar and is the maximum width of all intervals.\\
\end{comment}

%--------------------------------------------------------------
\subsubsection{\textbf{Set Operations and Arithmetic}}
%--------------------------------------------------------------

Every interval is a set, but not vice-versa \cite{Mustafa}. For a set to be an interval, it has to be closed, connected and defined in real numbers. Set operations include \emph{intersection}, \emph{union} and \emph{difference}. %For union, it can result in two disconnected intervals. \emph{Interval Hull} encompasses the entire area between two intervals that are disconnected, which is denoted by $\sqcup$. For example, let $[a]=[\underline{a},\overline{a}]$ and $[b] = [\underline{b},\overline{b}]$ then $[a]\cup[b] = [a]\sqcup[b] = [\mathrm{Min}(\underline{a},\underline{b}),\mathrm{Max}(\overline{a},\overline{b})]$. 
Set operations can also be extended to multiple dimensions.

Intervals also have arithmetic and functions that are applied extensively~\cite{Jaulin2001IntervalAnalysis}. That includes addition and multiplication at a basic level. Division and subtraction are best described as interval functions~\cite{Mustafa}.  
\begin{definition}[\textbf{Constraint Satisfaction Problem}]
A constraint satisfaction problem (CSP), in the context of interval analysis, is defined as the triple $\left(\mathbf{x},[\mathbf{x}],\mathbf{f}\left(\mathbf{x}\right) \leq \mathbf{0}\right)$, where $\mathbf{x} \in \mathbb{R}^n$ is a set of $n$ variables, $[\mathbf{x}] \in \mathbb{IR}^n$ is box representing the interval domain of each variable (cf. Definition~\ref{Interval}), and $\mathbf{f}\left(\mathbf{x}\right) \leq \mathbf{0}$ are the set of constraints such that $\mathbf{f}: \mathbb{R}^n \rightarrow \mathbb{R}^m$ is a vector function whose coordinates are $f_j$, $\forall j\in\{1,\dots,m\}$ \cite{Mustafa}. The solution set of the CSP is defined as:
\begin{equation}
    \label{eq:csp:CPS_solution}
	\mathbb{S}_{\textbf{x}} = \{\textbf{x} \in [\textbf{x}] \mid \textbf{f}(\textbf{x}) \leq \textbf{0}\}.
\end{equation}
\label{CSP}
\end{definition}

%--------------------------------------------------------------
\subsubsection{\textbf{Interval Methods}}
%--------------------------------------------------------------

\begin{figure}
    \centerline{
    \includegraphics[width=0.33\columnwidth]{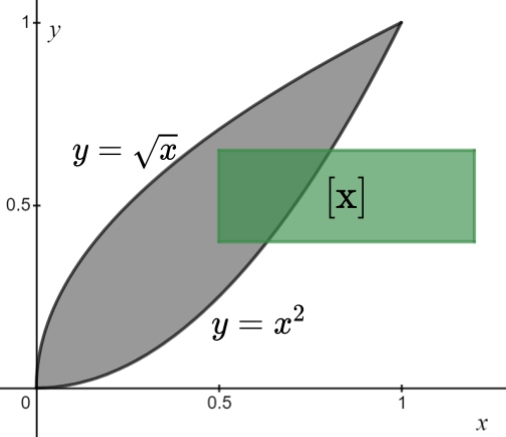}
    \includegraphics[width=0.33\columnwidth]{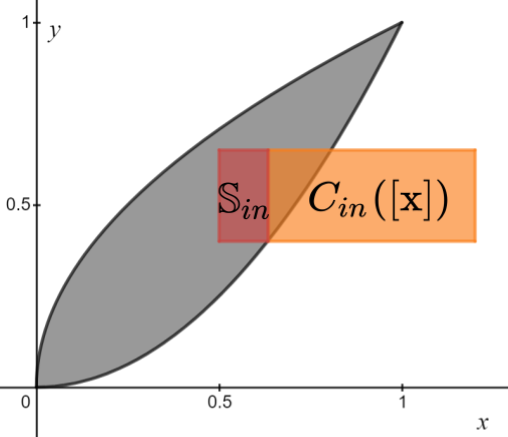}\includegraphics[width=0.33\columnwidth]{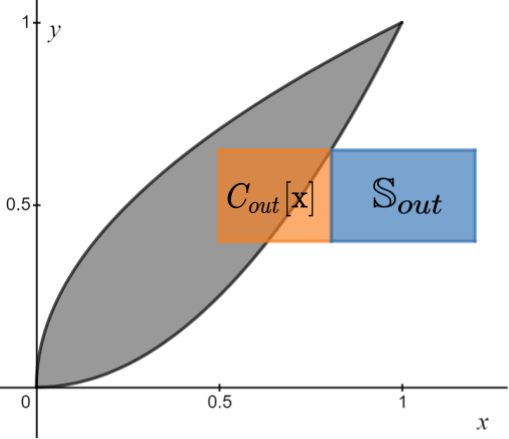}}
    \caption{CSP with $\mathrm{x} \in \mathbb{R}^2$, initial box $[\mathrm{x}] = [0.5, 1.2] \times [0.4, 0.6]$, and two constraints: $y \ge x^2\,,\, y \leq \sqrt{x}$. Inner contractor versus an outer contractor. $\mathbb{S}_{out}$ and $\mathbb{S}_{in}$ are the remainder of $\mathcal{C}_{out}$ and $\mathcal{C}_{in}$ actions~\cite{Mustafa}.}
    \label{fig:Contractors}
\end{figure}

Given a \emph{CSP}, interval methods are heuristic search algorithms that estimate an upper bound solution.
\begin{definition}[\textbf{\emph{Contractor}}]
\label{contractor}
 A \emph{contractor} is an interval method that estimates the solution of a given CSP (cf. Definition~\ref{CSP}) with the map $\mathcal{C}:\mathbb{R}^n\to\mathbb{R}^n$.
 Let $\mathbb{S}_{\mathrm{x}}$ be the set of solution and  $\mathcal{C}:\mathbb{R}^n\to\mathbb{R}^n$ be a contractor, which is applied to box $[\mathrm{x}]$ then $\mathcal{C}([\mathrm{x}])\subseteq [\mathrm{x}]$ and $\mathcal{C}([\mathrm{x}])\,\cap\, \mathbb{S}_{\mathrm{x}}=[\mathrm{x}]\,\cap\,\mathbb{S}_{\mathrm{x}}$. The former satisfies the \emph{contraction} condition and the latter satisfies \emph{correctness} condition \cite{Jaulin2001AppliedIA}. 
\end{definition}
As illustrated in Fig.~\ref{fig:Contractors}, we can have inner contractor $\mathcal{C}_{in}$ and outer contractor $\mathcal{C}_{out}$, which are complements of each other in terms of constraints.
%
\begin{comment}
\begin{equation}
\forall{[\mathrm{x}]\in \mathbb{IR}^n},\;\mathcal{C}([\mathrm{x}]) = [[\mathrm{x}]\cap\mathbb{S}].
    \label{eq:Minimal contractor}
\end{equation}
\end{comment}

After applying the contractor, the result is best described as three sets:
$\mathbb{S}_{in}$ is the estimated solution of $\mathbb{S}_{\mathrm{x}}$ such that $\mathbb{S}_{in} \subset \mathbb{S}_{\mathrm{x}}$ color-coded in red.
$\mathbb{S}_{out}$ is the \textit{complement} of the estimated solution such that $\mathbb{S}_{out} \cap \mathbb{S}_{\mathrm{x}} = \varnothing $ color-coded in blue.
$\mathbb{S}_{boundary}$ is the boundary set that represents the area that both include the solution and its inverse such that $\mathbb{S}_{boundary} \cap \mathbb{S}_{\mathrm{x}} \not = \varnothing$ and $\mathbb{S}_{boundary} \not\subset \mathbb{S}_{\mathrm{x}}$ color-coded in orange.

For different CSPs, there are many types of contractors\cite{hansen2003global,neumaier1990interval}. However, in this paper, we will use \emph{Forward-backward Contractor} \cite{Granvilliers1999RevisingConsistency} for its simplicity and efficiency.
\begin{definition}[\textbf{\emph{Forward-backward Contractor}}]
\label{Forward-backward Contractor} Forward-backward Contractor is a contractor (cf. Definition ~\ref{contractor}) that is applied to a CSP with one single constraint; it contracts in two steps: \emph{forward propagation} and \emph{backward propagation}~\cite{Mustafa}; they are illustrated in Algorithm~\ref{alg:fw}.
\end{definition}
\begin{center}
\begin{algorithm}
\caption{Forward-backward Contractor $\mathcal{C}_{\uparrow \downarrow}$.}
\label{alg:fw}
\begin{algorithmic}[1]
\WHILE {$\mathcal{C}_{\uparrow \downarrow}([\mathbf{x}],\; f(\mathbf{x}),\;[I])$}
\STATE {$[y] = [I] \cap [f]([\mathbf{x}])$}
\FORALL {$[x_i] \in \mathbf{[x]} : i\in\{1,...,n\}$}
\STATE {$[x_i]=[x_i] \cap [f^{-1}_{x_i}]([y],[\mathrm{x}])$}
\ENDFOR
\RETURN {$[\mathbf{x}]$}
\ENDWHILE 
\end{algorithmic}
\end{algorithm}
\end{center}

%\vspace{-1ex}
Generally, a CSP can have multiple constraints; to handle such a situation, we can use \emph{Outer Contractor}, which applies forward-backward contractor on each constraint independently as shown in Algorithm~\ref{alg:outer} \cite{Mustafa}.
\begin{center}
\begin{algorithm}
\caption{Outer Contractor $\mathcal{C}_{out}$.}
\label{alg:outer}
\begin{algorithmic}[1]
\WHILE {$\mathcal{C}_{out}([\textbf{x}],\; \textbf{f}(\textbf{x}))$}
\FORALL {$f_j \in \textbf{f} : j\in\{1,...,m\}$}
\STATE {$[\textbf{x}] \gets \mathcal{C}_{\uparrow \downarrow}([\textbf{x}],\;f_j(\textbf{x}),(-\infty,0])$}
\ENDFOR
\RETURN {$[\textbf{x}]$}
\ENDWHILE 
\end{algorithmic}
\end{algorithm}
\end{center}

%\vspace{-1ex}
Forward-backward contractor can be used to construct another type of contractor where the complementary CSP is considered; such contractor is called \emph{Inner Contractor}, and its procedure is defined in Algorithm~\ref{alg:inner} \cite{Mustafa}.

%\vspace{-1ex}
\begin{center}
\begin{algorithm}[h!]
\caption{Inner Contractor $\mathcal{C}_{in}$.}
\label{alg:inner}
\begin{algorithmic}[1]
\WHILE {$\mathcal{C}_{in}([\textbf{x}],\;\textbf{f}(\textbf{x}))$}
\STATE {$[\hat{\textbf{x}}] \gets \emptyset$}
\FORALL {$f_j \in \textbf{f} : j\in\{1,...,m\}$}
\STATE {$[\hat{\textbf{x}}] \gets[\hat{\textbf{x}}] \sqcup  \mathcal{C}_{\uparrow \downarrow}([\textbf{x}],\;f_j(\textbf{x}),(0,\infty))$}
\ENDFOR
\RETURN {$[\hat{\textbf{x}}]$}
\ENDWHILE
\end{algorithmic}
\end{algorithm}
\end{center}

\begin{comment}

Note that $f:\mathbb{R}^n \to \mathbb{R}$. (one constraint)
Example: $f(\mathbf{x}) = x_1^2 + \cos(x_2) - x_3$ where $n =3$,
 such that $\mathbf{x} = [x_1 \quad  x_2 \quad x_3]^T$

In the case of multiple constraints, $\mathbf{f} : \mathbb{R}^n \to \mathbb{R}^m$, such that: $f_1(\mathbf{x}) = x_1 + x_2 - \cos(x_3); f_2(\mathbf{x}) = x_1 - 2*x_2$

Note that the above two steps can be repeated iteratively until convergence is reached; this is known as a fixed-point contractor [ref].
\end{comment}

%----------------------------------------------------------------
\section{Incremental Symbolic Bounded Model Checking of Software Using Interval Methods via Contractors}
\label{sec:ProposedMethod}
%----------------------------------------------------------------

\begin{figure}[h!]
    \centering
\includegraphics[width=0.80\columnwidth]{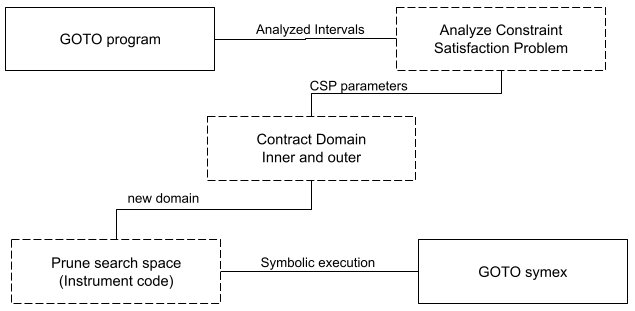}    
    \caption{Contractors in BMC.}
    \label{fig:Arch}
\end{figure}

Here, we introduce contractors described in Section~\ref{sec:IAM} to BMC of software. In particular, we model the constraints and properties generated from BMC instances to a CSP~\cite{tsang1993foundations}. In a CSP, we have three inputs: variables (dimensions), domain, and constraints. We obtain the domains by analyzing the declaration of independent variables, \textit{assume} statement and variable assignment (BMC constraints). The directive \textit{assert}, which represents the property, will give the contractor constraints. Assert statements will also dictate which variables are included in our defined domains. We apply the contractors in the intermediate representation (IR) of programs. Our IR is a GOTO-program that simplifies the program representation (e.g., replaces \textit{switch} and \textit{while} by \textit{if} and \textit{goto} statements)~\cite{DBLP:phd/ethos/Cordeiro11,Cordeiro2012SMT-basedSoftware}. Fig.~\ref{fig:Arch} shows how contractors can be used in BMC, where the \textit{GOTO-symex} component perform the program symbolic execution.

As an illustrative example, we consider the code fragment illustrated in Fig.~\ref{fig:code1}. We model this code into a CSP by having our variables  \texttt{x} as $x_1$ and \texttt{y} as $x_2$. Assume the maximum value that an unsigned integer can hold, defined as $\mathrm{Max}_{uint}$. Therefore, we define our intervals for each variable from the declaration of the variables. $[x_1]=[0,\mathrm{Max}_{uint}] , [x_2]=[0,\mathrm{Max}_{uint}]$. With the \textit{assume} directive, variable $x_1$ interval will be $[x_1]=[0,20]$. With the given assertion \texttt{x>=y}, we determine the inequality used being $x_1 \ge x_2$ as our constraint. Now we have our CSP, we model our input as parameters for a contractor such that: Variables $x_1,x_2$; Domains, $[x_1]=[0,20] , [x_2]=[0,\mathrm{Max}_{uint}]$; Constraint, $x_2-x_1 \leq 0$. 
\begin{figure}[ht]
    \centering
\begin{lstlisting}[language=c]
unsigned int x=nondet_uint();
unsigned int y=nondet_uint();
__ESBMC_assume(x <= 20);
assert(x >= y);
\end{lstlisting}
    \caption{Outer contractor example.}
    \label{fig:code1}
\end{figure}

 %Let $f(x) = x_2-x_1 $ such that:
%
\begin{figure}[ht]
\begin{center}
\fontsize{8}{8}\selectfont
\begin{tabular}{ ccc } 
$f(x) \leq 0$&$I = (-\infty,0]$ &  \\ [1ex]
$f(x): y = x_2-x_1$&$[y] = I \cap ([x_2] - [x_1])$&Forward-step\\[1ex]
$f(x)_{x_1}^{-1}:x_1 = x_2-y$&$[x_1]=[x_1] \cap ([x_2] - [y])$&Backward-step\\ [1ex]
$f(x)_{x_2}^{-1}:x_2 = y+x_1$&$[x_2] = [x_2] \cap ([y]+[x_1])$&Backward-step\\ [1ex]
\end{tabular}
\end{center}
\caption{Outer contractor steps for the example in Fig~\ref{fig:code1}.}
\label{fig:ocformula}
\end{figure}
\begin{figure}[ht]
    \centerline{
    \includegraphics[width=0.33\columnwidth]{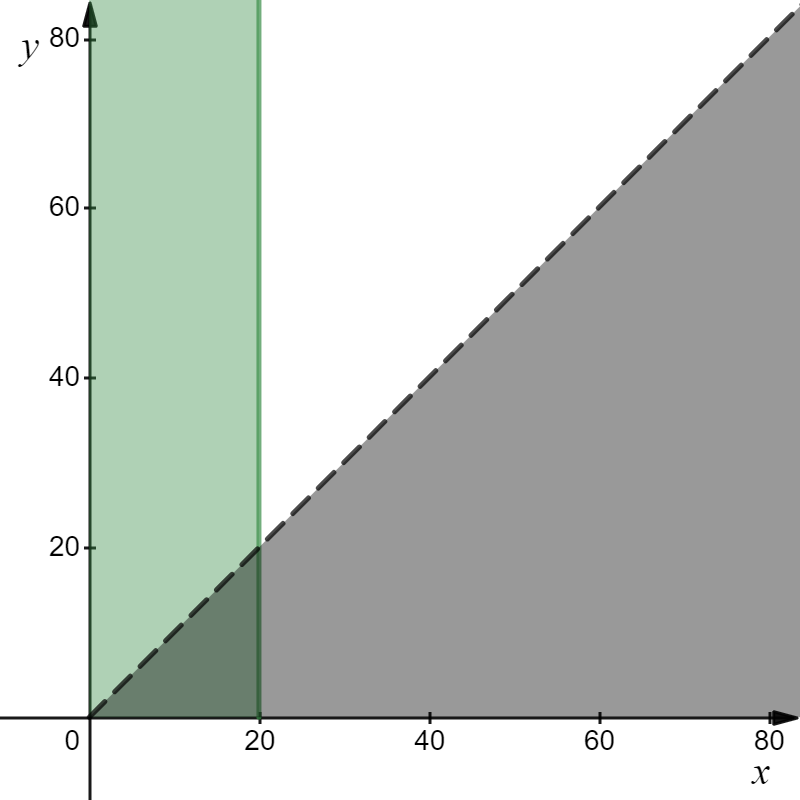} \includegraphics[width=0.33\columnwidth]{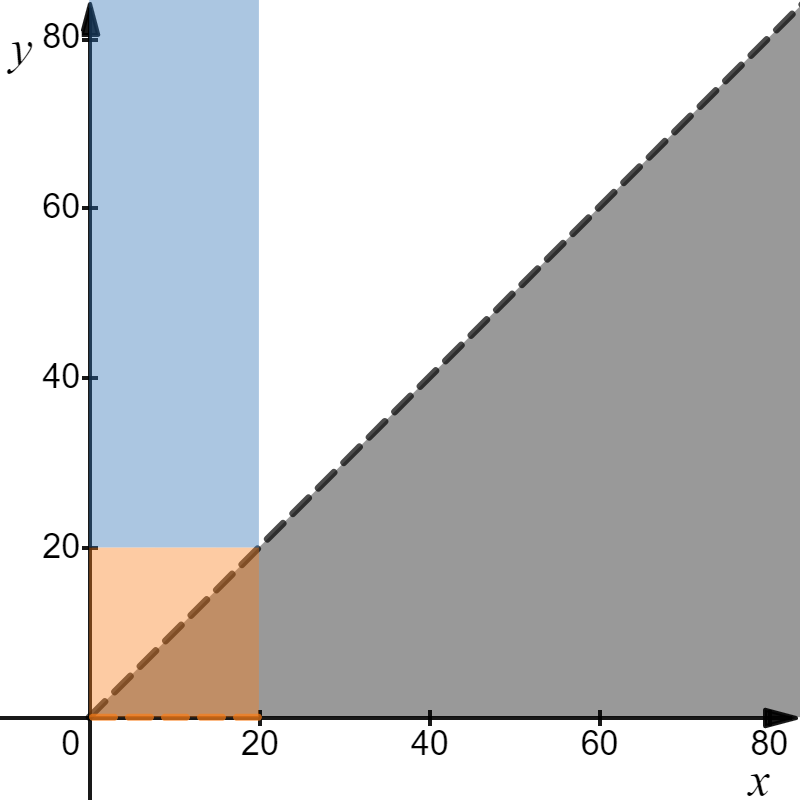}}
    \caption{Outer Contractors in BMC.}
    \label{fig:OContractorsBMC}
\end{figure}

We use the \textit{forward-backward} contractor~\cite{Granvilliers1999RevisingConsistency} mentioned in Definition~\ref{Forward-backward Contractor}. After we plug our values in the equations in Fig.~\ref{fig:ocformula}, we obtain $ [y] = [-20,0]$, $[x_1]=[0,20]$ and $[x_2] = [0,20]$. We notice that $[x_2]$ was contracted from $[0,\mathrm{Max}_{uint}]$ to $[0,20]$ because that is where our solution lies. Regarding BMC, having the domain contracted by the outer contractor ensures property violation since that area is outside the solution $\mathbb{S}_{out}$. We can visualize the results in Fig.~\ref{fig:OContractorsBMC}, where the left graph shows our initial domain, the right graph shows the contracted domain in blue, and the area remaining in orange. 

However, if we use the inner contractor, which is the complement of our constraint $f(x)>0$, we may prune the area inside our solution $\mathbb{S}_{in}$ to possibly prune our search space and only have the $\mathbb{S}_{boundary}$ area to be checked by BMC. Consider the example in Fig.~\ref{fig:code2}, we form our CSP by having our variables: let \texttt{x} be $x_1$ and \texttt{y} be $x_2$, and domain be $[x_1] = [20,30]$ and $[x_2] = [0,30]$. Constraint as $x_1\ge x_2$.\\
\begin{figure}[ht]
\centering
\begin{lstlisting}[language=c]
unsigned int x=nondet_uint();
unsigned int y=nondet_uint();
__ESBMC_assume(x >= 20 && x <= 30);
__ESBMC_assume(y <= 30);
assert(x >= y);
\end{lstlisting}
\caption{Inner contractor example.}
\label{fig:code2}
\end{figure}
\begin{figure}[ht!]
\begin{center}
\fontsize{8}{8}\selectfont
\begin{tabular}{ccc}
$f(x) > 0$&$I = [0,\infty)$ &  \\ [1ex]
$f(x):y = x_2-x_1$&$[y] = I \cap ([x_2] - [x_1])$&Forward-step\\ [1ex]
$f(x)_{x_1}^{-1}:x_1=x_2-y$&$[x_1]=[x_1] \cap ([x_2] - [y])$&Backward-step\\ [1ex]
$f(x)_{x_2}^{-1}:x_2 = y+x_1$&$[x_2] = [x_2] \cap ([y]+[x_1])$&Backward-step\\ [1ex]
\end{tabular}
\end{center}
\caption{Inner contractor steps for the example in Fig~\ref{fig:code2}.}
\label{fig:icformula}
\end{figure}

\begin{figure}[ht]
     \centerline{\includegraphics[width=0.33\columnwidth]{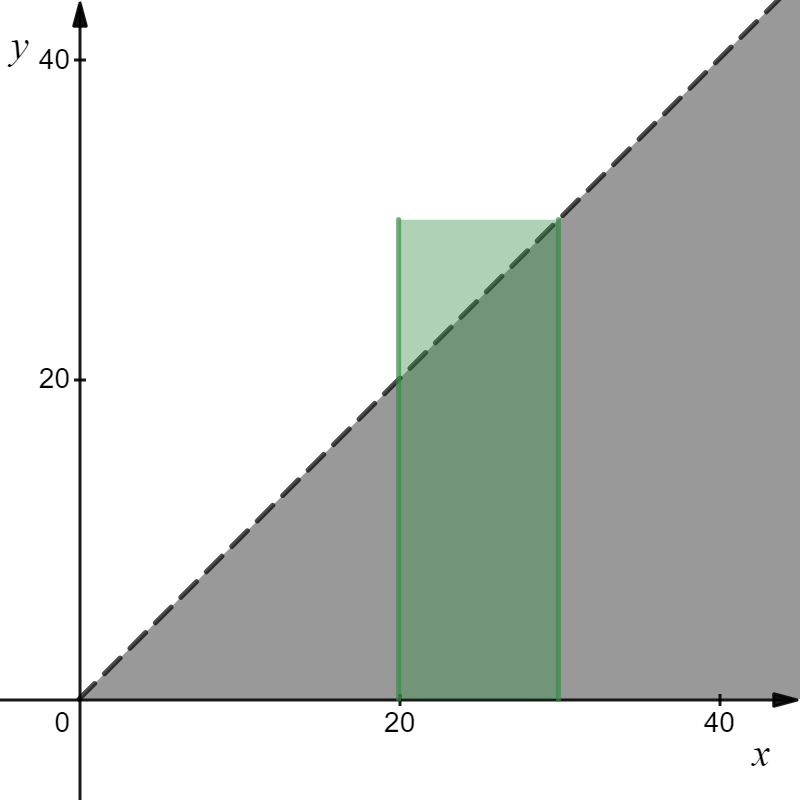}
    \includegraphics[width=0.33\columnwidth]{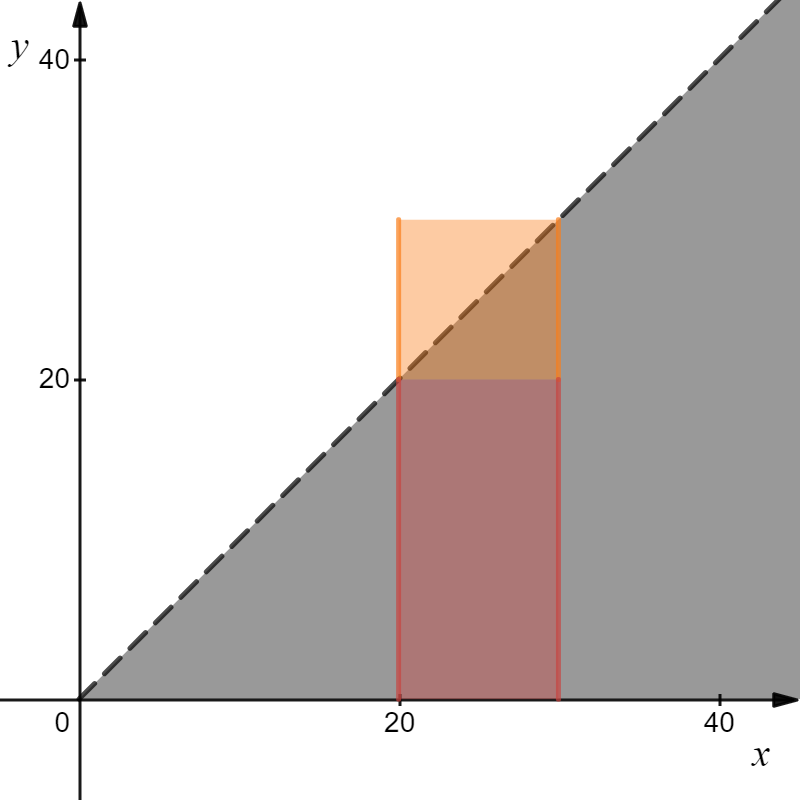}}
    \caption{Inner Contractors in BMC.}
    \label{fig:IContractorsBMC}
\end{figure}

For the inner contractor, we list our formulas as illustrated in Fig.~\ref{fig:icformula}. So we plug our values for $[x_1]$ and $[x_2]$; we obtain $[y] = [0,10]$, $[x_1] = [20,30]$ and $[x_2] = [20,30]$. Here, $[x_2]$ was contracted from $[0,30]$ to $[20,30]$. To visualize the solution, in Fig. \ref{fig:IContractorsBMC}, the red area is contracted from our domain; it represents the part of our domain that belongs to $\mathbb{S}_{in}$. For BMC, this area is guaranteed to hold the property; therefore, we prune our search space to the orange area $\mathbb{S}_{boundary}$ only. %

Algorithms~\ref{alg1} and~\ref{alg:contract} contain the main steps necessary for our approach. %
The first four lines describe each variable we will use throughout the process. We will analyze the variables $\textbf{x}$ and their intervals (or domain) $[\textbf{x}]$ and the properties (or constraints in the context of CSP) $\textbf{f}(\textbf{x}) \leq \textbf{0}$. Then, after calling $Apply\_Contractor$, we initialize the contractors based on variables and constraints. The outer contractors $\mathcal{C}_{out}$ are initialised based on the property in the form of $f(x) \leq 0$ or $f(x) = 0$, which depends on the property. In contrast, the inner contractor $\mathcal{C}_{in}$ is initialized in the form $f(x) > 0$ or $f(x) \neq 0$. 
\begin{algorithm}[htb!]
\caption{Proposed method algorithm.}
 \label{alg1}
 \centering
 \begin{algorithmic}[1]
 \REQUIRE {Program $P$}
 \STATE {$\textbf{x}$ : set of Variables}
 \STATE {$[\textbf{x}]$ : domain}
 \STATE {$\textbf{f}(\textbf{x}) \leq \textbf{0}$ : set of Constraints}
\STATE {$\textbf{f}(\textbf{x}) \leq \textbf{0} \gets Analyze\_Properties(P)$}
 \STATE {$[\textbf{x}] \gets Analyze\_Intervals(P)$}
 \STATE {$[\textbf{x}] \gets Apply\_Contractor(\textbf{x},[\textbf{x}],\textbf{f}(\textbf{x}) \leq \textbf{0})$}
 \STATE $P^\prime \gets instrument(P,[\textbf{x}])$
\RETURN {$P^\prime$}
 \end{algorithmic}
\end{algorithm}

\begin{algorithm}[htb!]
\caption{$Apply\_Contractor$.}
 \label{alg:contract}
 \centering
 \begin{algorithmic}[1]
 \REQUIRE {Varibles $\textbf{x}$, Domains $[\textbf{x}]$, Constraints $\textbf{f}(\textbf{x})\leq 0$}
 \STATE {$[\textbf{x}]^\prime \gets \mathcal{C}_{out}([\textbf{x}],\;\textbf{f}(\textbf{x}) )$} 
 \IF {$[\textbf{x}]^\prime = \emptyset$}\label{lst:line:2}
 \STATE $P^\prime \gets P$\label{lst:line:3}
 \ELSE
 \STATE $\mathbb{S}_{out} \gets [\textbf{x}] \setminus [\textbf{x}]^\prime$
 \STATE $[\textbf{x}]^{\prime\prime} \gets \mathcal{C}_{in}([\textbf{x}]^\prime,\;\textbf{f}(\textbf{x}))$
 \STATE $\mathbb{S}_{in} \gets [\textbf{x}]^\prime \setminus [\textbf{x}]^{\prime\prime}$
 \STATE $\mathbb{S}_{boundary} \gets [\textbf{x}]^{\prime\prime}$
 \ENDIF
 \RETURN {$\mathbb{S}_{boundary} \cup \mathbb{S}_{out}$}
 \end{algorithmic}
 \end{algorithm}

With CSP components being set up, we contract our domain starting with $\mathcal{C}_{out}$, where the domain will be reduced to exclude the region outside the solution, which violates the property. If the resulting set of the domain $[\textbf{x}]^\prime$ is empty, it means that the domain will violate the property(s). There is no need to prune the search space (lines \ref{lst:line:2} and \ref{lst:line:3} of Algorithm~\ref{alg:contract}). Otherwise, the difference between the result from the outer contractor $[\textbf{x}]^\prime$ and the original domain $[\textbf{x}]$ will yield $\mathbb{S}_{out}$, as demonstrated in line $10$ of Algorithm~\ref{alg1}. After that, we apply the inner contractor $\mathcal{C}_{in}$ to remove the intervals inside the solution where the property holds. Similar to the outer contractors, the difference between the two sets $[\textbf{x}]^{\prime\prime}$ and $[\textbf{x}]^{\prime}$ will yield the set inside the solution $\mathbb{S}_{in}$. Note that box difference will not necessarily yield one single box but rather a finite set of boxes \cite{Mustafa}. The resulting domain from the contractor $[\textbf{x}]^{\prime\prime}$ will be our boundary area, where $[\textbf{x}]^{\prime\prime} \subseteq \mathbb{S}_{boundary}$. With $\mathbb{S}_{out}$, $\mathbb{S}_{in}$ and $\mathbb{S}_{boundary}$, now we can instrument the program for the BMC engine to search for a counterexample in $\mathbb{S}_{out}$ and $\mathbb{S}_{boundary}$ while removing $\mathbb{S}_{in}$ from the search-space. 

The instrumentation will be done using the \texttt{assume(expr)} directive, where we constrain the search space indicated by the expression $expr$, which the contractors produced. Note that instrumenting \texttt{assume(expr)} directive is not a trivial task since we need to ensure where the pruned intervals are in the context of the program, especially when it comes to loops. If $\mathbb{S}_{in}$ is at the end of the loop, we add the \texttt{assume($\mathbb{S}_{out}\cup \mathbb{S}_{boundary}$)} to remove these unnecessary loop steps in the loop. However, if $\mathbb{S}_{in}$ is at the beginning or middle of the loop, we discussed it further in Section~\ref{sec:TTTV}.

\begin{figure}[ht]
    \centerline{
    \includegraphics[width=0.33\columnwidth]{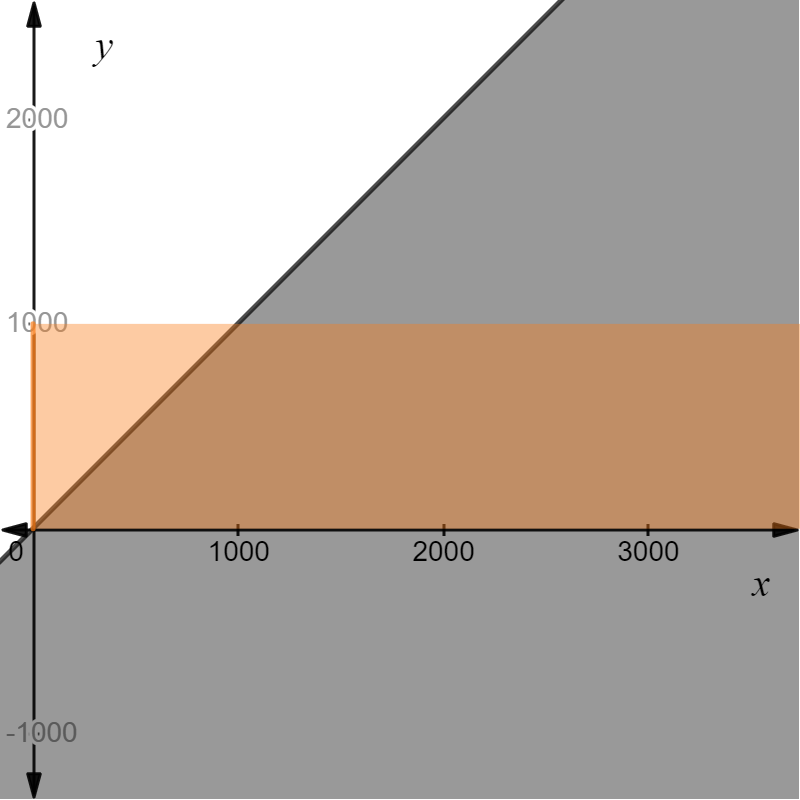} \includegraphics[width=0.33\columnwidth]{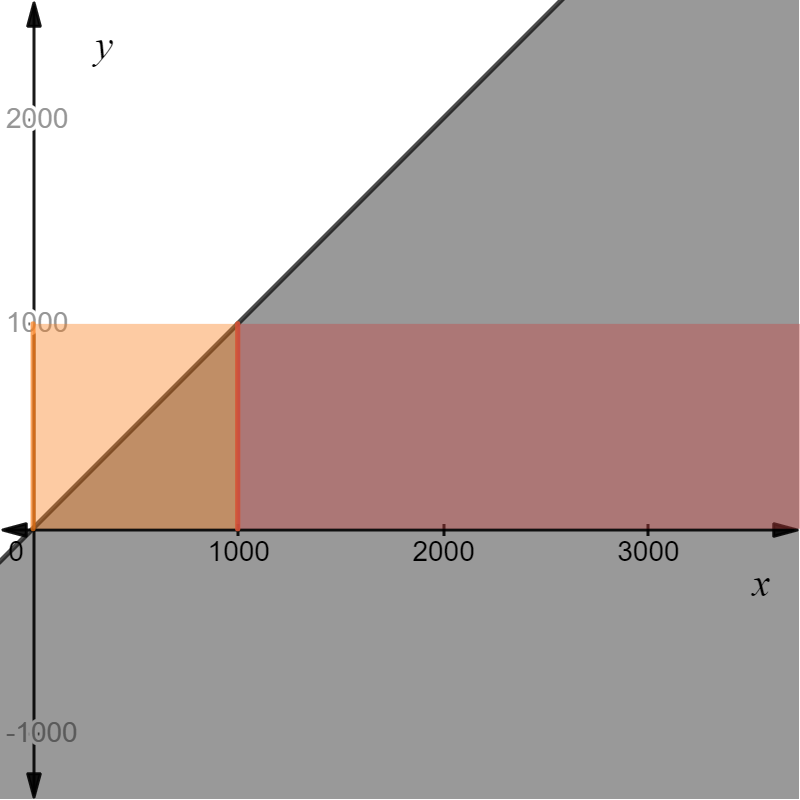}}
    \caption{\textit{afnp2014.c} search space before and after contractors.}
    \label{fig:afnp_graph}
\end{figure}

We will use the illustrative example in Fig.~\ref{fig:exp1} to demonstrate the application of the steps of Algorithm~\ref{alg1}. Note that we have two variables in this program \texttt{x} and \texttt{y}, which will be $x_1$ and $x_2$ in the CSP space, respectively, where $\textbf{x}=\{x_1,x_2\}$.
For the domain, after the analysis, they will be $x_1 = [1,\mathrm{Max}_{int}]$ and $x_2 = [0,1000]$ with $[\textbf{x}]=[1,\mathrm{Max}_{int}]\times[0,1000]$, while the constraint will be $x_1 \ge x_2$ where $\textbf{f}(\textbf{x}) \leq \textbf{0}\Rightarrow f(x)=x_2 - x_1 \leq 0$. Therefore, our constraints will be in the form of $x_2-x_1\leq 0$ for the outer contractor and $x_2-x_1 > 0$ for the inner contractor.  In this case, however, the outer contractor returned the same domain, which means it cannot be contracted from the outside because there are no values in the domain $[\textbf{x}]$ guaranteed to be outside the solution, thus $\mathbb{S}_{out}=\emptyset$. Fig.~\ref{fig:afnp_graph} illustrates the outcome of the inner contractor on the program (line 12 of Algorithm~\ref{alg1}), where the domain were reduced from $[\textbf{x}]^\prime=[1,\mathrm{Max}_{int}]\times[0,1000]$ to $[\textbf{x}]^{\prime\prime}=[1,1000]\times[1,1000]$ where $\mathbb{S}_{in} = \{[1000,\mathrm{Max}_{int}]\times[0,1000] , [0,1000]\times[0,1]\}$ and $\mathbb{S}_{boundary}\gets [\textbf{x}]^{\prime\prime}$. 
An example of the employed instrumentation is shown in Fig.~\ref{fig:afnp2014.cWithContractor}. The \texttt{assume(expr)} directive here is placed in the loop, where the values change for variables. It is placed in this location to guide the verifier not to check steps further in the loop because these steps represent $\mathbb{S}_{in}$, where the property is guaranteed to hold. 

\begin{figure}[ht]
    \centering
    \begin{lstlisting}[language=c]
#include <assert.h>
int main() {
    int x = 1, y = 0;
    while (y < 1000 
            && __VERIFIER_nondet_int()) {
        x = x + y;
        y = y + 1;
    }
    assert(x >= y);
    return 0;
}
\end{lstlisting}
    \caption{C program extracted from \textit{sv-benchmarks/c/loop-lit/afnp2014.c}, where \cou{\_\_VERIFIER\_nondet\_int$()$} returns a non-deterministic integer.}
    \label{fig:exp1}
\end{figure}
\begin{figure}[ht]
    \centering
    \begin{lstlisting}[language=c]
#include <assert.h>
int main() {
    int x = 1, y = 0;
    while (y < 1000 
            && __VERIFIER_nondet_int()) {
        x = x + y;
        y = y + 1;
        assume(x <= 1000);
    }
    assert(x >= y);
    return 0;
}
\end{lstlisting}
    \caption{\textit{afnp2014.c} after applying contractors.}
    \label{fig:afnp2014.cWithContractor}
\end{figure}

%---------------------------------------------------------------- 
\section{Implementation}
%----------------------------------------------------------------

The implementation of the proposed approach follows the Algorithm~\ref{alg1} and goes into the following steps, as described below. Though it has limitations, we will discuss them further in Section~\ref{lim}.

%----------------------------------------------------------------
\subsection{Property Analysis}
%----------------------------------------------------------------

In this step, we will analyze the property used as a constraint in the CSP. We start with parsing each expression in the \cou{assert} directive, which is carried by converting from the syntax of ESBMC GOTO-program to the IBEX syntax. Figures~\ref{fig:ESBMC_Syntax}and~\ref{fig:IBEX_Syntax} illustrate both syntaxes, where we notice that IBEX syntax is more strict in its structure than ESBMC's syntax. For example, $(x>1)+1$ is allowed in ESBMC's syntax but not IBEX. Furthermore, IBEX syntax does not have an operator for $!=$, which makes it hard to convert some expressions. While parsing an expression, we build a list of variables inside the assertions, which we will use later to analyze intervals. We have two lists updated from this procedure: a list of variables and a list of constraints. 
\begin{figure}
\scriptsize
    \centering
    \begin{tabular}{c c l l}
    $n$ & $\in$ & $\mathbb{V}$ & scalar values\\
    $\cou{x}$ & $\in$ & $\mathbb{X}$ & program variable\\
    $\odot $ & $\Coloneqq$ & $+\,|\,-\,|\,*\,|\,...$ & binary operator\\
    $\circledless $ & $\Coloneqq$ & $<\,|\,\leq\,|\,==\,|\,...$ & comparison operator\\
    $\expression $ & $\Coloneqq$ & $ $ & scalar expressions\\
    $ $ & $|$ & $n$ & scalar constant\\
    $ $ & $|$ & $\cou{x}$ & variable\\
    $ $ & $|$ & $\expression\;\odot\;\expression$ & binary operation\\
    $\bexpression$ & $\Coloneqq$ & $ $ & Boolean expressions\\
    $ $ & $|$ & $\expression\;\circledless\;\expression$ & Compare two expressions\\
    $\logic$ & $\Coloneqq$ & $ $& logical expression\\
    $ $ & $|$ & $\bexpression$ & \\
    $ $ & $|$ & $\bexpression \wedge \bexpression$ & and\\
    $ $ & $|$ & $\bexpression \vee \bexpression$ & or\\
\end{tabular}
    \caption{IBEX Syntax.}
    \label{fig:IBEX_Syntax}
\end{figure}
\begin{figure}
\scriptsize
    \centering
   \begin{tabular}{c c l l}
    %$n$ & $\in$ & $\mathbb{V}$ & scalar values\\
    %$\cou{x}$ & $\in$ & $\mathbb{X}$ & program variable\\
    %$\odot $ & $\Coloneqq$ & $+\,|\,-\,|\,*\,|\,...$ & binary operator\\
    %$\circledless $ & $\Coloneqq$ & $<\,|\,\leq\,|\,==\,|\,...$ & comparison operator\\
    $\expression $ & $\Coloneqq$ & $ $ & scalar expressions\\
    $ $ & $|$ & $n$ & scalar constant\\
    $ $ & $|$ & $\cou{x}$ & variable\\
    $ $ & $|$ & $\expression\;\odot\;\expression$ & binary operation\\
    $ $ & $|$ & $\expression\;\circledless\;\expression$ & Compare two expressions\\
    $ $ & $|$ & $\expression \wedge \expression $ & and\\
    $ $ & $|$ & $\expression \vee \expression $ & or\\
    $ $ & $|$ & $\neg \expression $ & not\\
    $ $ & $|$ & $Typecast(\expression) $& typecast\\
    ...
\end{tabular}
    \caption{ESBMC goto-program Syntax.}
    \label{fig:ESBMC_Syntax}
\end{figure}

%-------------------------------------------------
\subsection{Interval Analysis}
%-------------------------------------------------

Interval analysis builds intervals for each variable that is in the assert statements. These intervals will serve as domains for the CSP. The current implementation now utilizes \texttt{--interval-analysis} option in ESBMC and parses each assume statement inserted by it. %The parsing is not as complex as in the previous section due to the form they are is much simpler. Usually they are in the form of \texttt{(<var\_name> <cop> <value>)}, where \texttt{var\_name} is the variable name and \texttt{<cop>} is a comparison operator. 
At each assume, we check whether the variable used is in our list, and if it is, we update its interval. 

%-------------------------------------------------
\subsection{Interval Methods (Apply Contractor)}
%-------------------------------------------------

For contractors, we chose IBEX library.\footnote{\url{ http://www.ibex-lib.org/}} IBEX is a C++ library for constraint processing over real numbers. It started in $2007$ as an open-source academic project that provides algorithms for handling non-linear constraints. IBEX is also the only Interval Analysis library that implements contractors. 

The contractor type used is the forward-backward contractor. With the constraint and the variables parsed from the first step, and domains from the second step, we create the contractor with the constraint complement to reduce the intervals inside the solution. As described previously, the contractor will need a CSP; at this point, we should have all the components to apply it. We have the list of constraints, the variables, and their domains. The resulting domains from the contractor will represent where the property may be violated, thus limiting the search space for BMC. 

%-------------------------------------------------
\subsection{Modify Original Program}
%-------------------------------------------------

With the resulting intervals, we compare them to the original and insert the \cou{assume} directive based on the difference. This is to tell BMC to ignore the instructions (loop iterations), which will result in holding the property. 

%-------------------------------------------------
\subsection{limitations}
\label{lim}
%-------------------------------------------------

In property analysis, we only support a limited number of conversions, which are one-to-one conversions. However, some expressions require a more complex conversion. For example, IBEX does not support the \texttt{!=} operator since a union of intervals can represent it. Another example is \texttt{(x>1)+1}, where it is possible to have in GOTO-program syntax but not in IBEX's syntax. That is because IBEX syntax would not allow adding a logical expression to a scalar or any other expression, as illustrated in Fig.~\ref{fig:IBEX_Syntax}. So the conversion of such expressions will be more complex and not a one-to-one conversion. Also, we only deal with one constraint per time. Additionally, we will not convert expressions if the property has an unsupported operator, which includes $!=$, $==$, $\&\&$, $||$.

The interval analysis option's drawback in ESBMC is that the intervals are sometimes too broad to be helpful for the contractor. Additionally, the lack of floating-point interval analysis could hinder the future since IBEX deals exclusively with \texttt{double} datatype.

%----------------------------------------------------------------
\section{Experimental Evaluation}
\label{sec:ExperimentalEvaluation}
%----------------------------------------------------------------

\begin{table*}[ht!]
\caption{Results for each subcategory. Each cell in the table shows two values. The top value represents the score, while the bottom represents time in seconds.}
\label{Table1}
\centering
\def\arraystretch{1.25}
\begin{tabular}{|c|c|c|c|c|c|c|c|}
\hline &&&&&&&\\%[-1.5ex]
\multirow{3}{*}{\textbf{Subcategory} / $\frac{Score}{Time(s)}$}&  \textbf{ESBMC}  &\textbf{ESBMC}&\textbf{ESBMC}&\multirow{2}{*}{\textbf{CBMC }}&\multirow{2}{*}{\textbf{Veriabs }}&\textbf{CPA}&\multirow{2}{*}{\textbf{Verifuzz}}\\
 & incremental-BMC &  incremental-BMC & k-induction &&&\textbf{Checker}&\\
 &with contractors&v6.9&v6.8&v5.43.0&v1.4.2&v2.1&v1.2.11\\
\hline
\multirow{2}{*}{\textbf{Arrays}}&159&151&234&-66&\textbf{729}&77&109\\&272607&273123&224050&133708&17449&352273&274219\\\hline
\multirow{2}{*}{\textbf{BitVectors}}&24&55&55&41&\textbf{76}&71&15\\&13876&14145&12924&8294&6077&9680&28691\\\hline
\multirow{2}{*}{\textbf{ControlFlow}}&58&55&64&67&\textbf{128}&113&40\\&47384&46727&23883&30696&24763&13987&46734\\\hline
\multirow{2}{*}{\textbf{ECA}}&270&268&1098&170&\textbf{1312}&930&429\\&947500&949801&539341&989186&449183&687361&799332\\\hline
\multirow{2}{*}{\textbf{Floats}}&800&802&779&782&\textbf{827}&749&-64\\&48147&46768&36501&46743&44156&62876&368963\\\hline
\multirow{2}{*}{\textbf{Heap}}&256&227&226&\textbf{292}&204&282&57\\&55180&53887&45732&48786&40801&50371&161308\\\hline
\multirow{2}{*}{\textbf{Loops}}&669&652&609&637&\textbf{1044}&696&209\\&296102&294764&273893&236902&170916&292803&492657\\\hline
\multirow{2}{*}{\textbf{ProductLines}}&-595&-595&783&256&903&\textbf{903}&216\\&214485&214554&62777&128144&149945&80606&48716\\\hline
\multirow{2}{*}{\textbf{Recursive}}&-64&-64&109&100&\textbf{125}&92&43\\&31790&31771&22736&18695&22073&23215&54177\\\hline
\multirow{2}{*}{\textbf{XCSP}}&\textbf{159}&159&158&158&152&153&55\\&14306&14334&13273&12087&19869&20442&59186\\\hline
\multirow{2}{*}{\textbf{AWS-C-Common}}&\textbf{174}&174&157&6&-&43&0\\&32234&33479&19637&5279&-&81193&1999\\\hline
\multirow{2}{*}{\textbf{BusyBox}}&0&0&2&0&-&\textbf{2}&0\\&2141&1812&10&7755&-&2555&206\\\hline
\multirow{2}{*}{\textbf{DeviceDriversLinux64}}&66&66&2020&-3257&-&\textbf{3114}&0\\&2133570&2102103&802304&1697560&-&948362&328643\\\hline
\multirow{2}{*}{\textbf{total}}&1976&1950&6294&-814&5500&\textbf{7225}&1109\\&4109322&4077268&2077061&3363835&945232&2625724&2664831\\\hline
\end{tabular}
\end{table*}

%\textcolor{blue}{In order to evaluate the proposed method, we conducted a set of experiments to compare the improvements of the BMC engine after applying contractors.}

%----------------------------------------------------------------
\subsection{Objectives}
\label{sec:Objectives}
%----------------------------------------------------------------

This experiment will compare results from performing incremental-BMC (cf. Definition~\ref{def:incrBMC}) with and without contractors in terms of time and memory consumption. This evaluation will answer the following main experimental goals: 
\begin{tcolorbox}
\begin{enumerate}
\item[EG1] \textbf{(Efficiency)} Does the contractors prune the search space to consume less resources when verifying programs with ESBMC?
\item[EG2] \textbf{(Soundness and completeness)} Do the contractors and resulting code instrumentation preserve the soundness and completeness when verifying programs with ESBMC?
\item[EG3] \textbf{(Improve ESBMC in programs with large loops)} Does our approach improve ESBMC performance in programs with large loops?
\end{enumerate}
\end{tcolorbox}

%----------------------------------------------------------------
\subsection{Description of the Setup}
\label{sec:BenchmarksandSetup}
%----------------------------------------------------------------

%We chose ESBMC~\cite{esbmc2018} as our BMC engine due to its incremental algorithms and its integration with benchexec. 
We executed ESBMC with the  same set of options of SV-COMP 2022, differing only on the strategy chosen (i.e., \textit{k}-induction or incremental-bmc).
%following set of options:  \texttt{--incremental-bmc}, which enables the incremental BMC; \texttt{--unlimited-k-steps}, which removes the upper limit of iteration steps in the incremental BMC algorithm; \texttt{--goto-contractor} to enable our method. 
%Experimental results that do not use the contractor were taken from SV-COMP 2022\footnote{\url{https://sv-comp.sosy-lab.org/2022/results/results-verified/}}. 
%Note that ESBMC supports the SMT floating-point logic QF\_ABVFP by default~\cite{GadelhaCN20}. 
We compare our results with state-of-the-art tools from SV-COMP 2022. These tools include CBMC~\cite{kroening2014cbmc} for its success over the years. Veriabs~\cite{darke2021veriabs}, which is the SV-COMP 2022 champion in this category and the runner-up CPA Checker 2.1~\cite{beyer2011cpachecker} and Verifuzz~\cite{chowdhury2019verifuzz}. SV-COMP score system consists in: (+2) for correct true, (+1) for correct false, (-32) for incorrect true, and (-16) for incorrect false.

All experiments were conducted on an Intel(R) Xeon(R) CPU E5-2620 v4 @ 2.10GHz and 160GB of RAM. We set time and memory limits of $900$ seconds and $15$GB for each benchmark, respectively. Finally, we present the results by showing the verification time and scores. All presented execution times are CPU times, i.e., only the elapsed periods spent in the allocated CPUs. Furthermore, memory consumption is the amount of memory that belongs to the verification process and is currently present in RAM (i.e., not swapped or otherwise not-resident). CPU time and memory consumption were measured with the $benchexec$ tool~\cite{beyer2016reliable} (commands available on the supplementary page). We did not enable swapping or turbo during our experiments, and all executed tools were restricted to a single core.

%----------------------------------------------------------------
\subsection{Description of the Benchmarks}
\label{sec:Benchmarks}
%----------------------------------------------------------------

We evaluate our approach using 7044 tasks in \texttt{unreach-call} benchmarks from SV-COMP $2022$~\cite{sv-comp22}, which can be described as \emph{verifying whether exists an execution path that can lead a benchmark to an assertion failure}. This category contains benchmarks extracted from various domains (e.g., linux drivers, recursion, unbounded loops, algorithms, real-world, etc.). Due to massive arithmetic computation within loops, this category can result in many timeouts (i.e. a tool can not verify it in a given time constraint); they fit our approach. All tools, benchmarks, and evaluation results are available on a supplementary web page\footnote{removed for double-blind review.}.  %\url{https://doi.org/10.5281/zenodo.6949341}}.

%We evaluate our approach using various benchmarks from SV-COMP $2022$~\cite{sv-comp22}. In particular, some of these benchmarks can be described as time-consuming, i.e., reaching the result as time-out due to massive arithmetic computation within loops. All the benchmarks fall under the \emph{unreach-call} property; they fit with our approach as they have asserts and lengthy loops. This \emph{unreach-call} property  \textcolor{blue}{can be described as whether exists an execution path that can lead a benchmark to reach an assertion failure.} % is best describedas: \cou{CHECK( init(main()), LTL(G ! call(func())) )} where \cou{init(main())} is the initial states of the program when we  call the function ``main''. The LTL (linear-time temporal logic) \cou{LTL(f)} specifies that formula \cou{f} holds at every initial state . Then, operator \cou{G f} means that formula \cou{f} holds globally. If the function \cou{func()} is called, the proposition \cou{call(func())} is true. Lastly, The \cou{!} operator is to make sure that function \cou{func()} is never triggered. 
%All tools, benchmarks, and evaluation results are available on a supplementary web page\footnote{removed for double-blind review.}  %\url{https://doi.org/10.5281/zenodo.6949341}}.

%----------------------------------------------------------------
\subsection{Results}
\label{sec:Results}
%----------------------------------------------------------------

\begin{table*}[ht!]
\caption{Results for loops category. Each cell in the table shows two values. The top value represents the score, while the bottom represents time in seconds.}
\label{Table2}
\centering
\def\arraystretch{1.25}
\begin{tabular}{|c|c|c|c|c|c|c|c|}
\hline &&&&&&&\\%[-1.5ex]
\multirow{3}{*}{\textbf{Subcategory} / $\frac{Score}{Time(s)}$}&  \textbf{ESBMC}  &\textbf{ESBMC}&\textbf{ESBMC}&\multirow{2}{*}{\textbf{CBMC }}&\multirow{2}{*}{\textbf{Veriabs }}&\textbf{CPA}&\multirow{2}{*}{\textbf{Verifuzz}}\\
 & incremental-BMC &  incremental-BMC & k-induction &&&\textbf{Checker}&\\
 &with contractors&v6.9&v6.8&v5.43.0&v1.4.2&v2.1&v1.2.11\\
\hline
\multirow{2}{*}{\textbf{loops}}&29&25&1&52&\textbf{85}&78&28\\&18353&18544&13261&14172&5817&10579&32777\\\hline
\multirow{2}{*}{\textbf{loop-acceleration}}&26&29&28&28&\textbf{49}&32&17\\&18568&15803&13734&5800&2967&16468&18698\\\hline
\multirow{2}{*}{\textbf{loop-crafted}}&3&3&8&1&\textbf{11}&5&0\\&3846&3608&1723&1450&459&3306&5403\\\hline
\multirow{2}{*}{\textbf{loop-invgen}}&11&11&15&3&\textbf{41}&37&1\\&18610&17661&15683&21295&8529&12144&25219\\\hline
\multirow{2}{*}{\textbf{loop-lit}}&15&15&13&23&\textbf{37}&21&1\\&12628&12645&11945&10011&4317&11150&18914\\\hline
\multirow{2}{*}{\textbf{loop-new}}&8&4&12&12&\textbf{16}&10&0\\&6309&8354&6285&3996&2835&6977&9906\\\hline
\multirow{2}{*}{\textbf{loop-industry-pattern}}&16&16&28&16&32&\textbf{36}&0\\&9061&9061&2923&8137&2363&1826&16208\\\hline
\multirow{2}{*}{\textbf{loops-crafted-1}}&-24&-28&-17&3&\textbf{96}&9&8\\&42493&40099&19040&27417&6344&42570&39663\\\hline
\multirow{2}{*}{\textbf{loop-invariants}}&3&3&3&1&\textbf{17}&7&1\\&7124&7122&6950&6141&144&4914&7208\\\hline
\multirow{2}{*}{\textbf{loop-simple}}&\textbf{13}&13&11&9&11&7&1\\&961&920&903&605&1001&2419&6307\\\hline
\multirow{2}{*}{\textbf{loop-zilu}}&20&14&16&14&\textbf{42}&38&0\\&10932&13657&12650&13135&1194&4887&19812\\\hline
\multirow{2}{*}{\textbf{verifythis}}&\textbf{2}&2&2&2&0&-30&2\\&3631&3634&3618&2069&2392&1939&2883\\\hline
\multirow{2}{*}{\textbf{nla-digbench}}&7&7&8&5&\textbf{23}&5&4\\&23947&23953&23617&21905&16700&25427&24412\\\hline
\multirow{2}{*}{\textbf{nla-digbench-scaling}}&540&538&481&468&\textbf{584}&441&146\\&119639&119702&141561&100768&115853&148198&265249\\\hline
\multirow{2}{*}{\textbf{total}}&669&652&609&637&\textbf{1044}&696&209\\&296102&294764&273893&236902&170916&292803&492657\\\hline
\end{tabular}
\end{table*}

\begin{table}[ht]
\caption{Results per category with memory consumption. Each cell shows the score at the top and the memory used in MB at the bottom. The decrease in memory usage is shown as the percentage of how much less memory was used. }
\centering
\label{table3}
\def\arraystretch{1.25}
\begin{tabular}{|c|c|c|c|}
\hline &&&\\%[-1.5ex]
\multirow{3}{*}{\textbf{Subcategory}}&  \textbf{ESBMC }  &\textbf{ESBMC }&\textbf{memory }\\
 &  incremental & incremental & \textbf{saved} \\
&with contractors&v6.9&\textbf{}\\
\hline
\multirow{2}{*}{\textbf{Arrays}}&159&151&\multirow{2}{*}{78\%}\\&152018&682100&\\\hline
\multirow{2}{*}{\textbf{BitVectors}}&24&55&\multirow{2}{*}{-10\%}\\&5877&5325&\\\hline
\multirow{2}{*}{\textbf{ControlFlow}}&58&55&\multirow{2}{*}{-23\%}\\&129040&104705&\\\hline
\multirow{2}{*}{\textbf{ECA}}&270&268&\multirow{2}{*}{-2\%}\\&3662494&3595599&\\\hline
\multirow{2}{*}{\textbf{Floats}}&800&802&\multirow{2}{*}{59\%}\\&74664&181319&\\\hline
\multirow{2}{*}{\textbf{Heap}}&256&227&\multirow{2}{*}{-16\%}\\&71968&62046&\\\hline
\multirow{2}{*}{\textbf{Loops}}&669&652&\multirow{2}{*}{36\%}\\&177184&275648&\\\hline
\multirow{2}{*}{\textbf{ProductLines}}&-595&-595&\multirow{2}{*}{4\%}\\&136344&141732&\\\hline
\multirow{2}{*}{\textbf{Recursive}}&-64&-64&\multirow{2}{*}{-9\%}\\&76335&69961&\\\hline
\multirow{2}{*}{\textbf{XCSP}}&159&159&\multirow{2}{*}{58\%}\\&12773&30461&\\\hline
\multirow{2}{*}{\textbf{AWS-C-Common}}&174&174&\multirow{2}{*}{89\%}\\&68463&622038&\\\hline
\multirow{2}{*}{\textbf{BusyBox}}&0&0&\multirow{2}{*}{96\%}\\&2216&54352&\\\hline
\multirow{2}{*}{\textbf{\scriptsize DeviceDriversLinux64}}&66&66&\multirow{2}{*}{83\%}\\&7045822&40824348&\\\hline
\multirow{2}{*}{\textbf{total}}&1976&1950&\multirow{2}{*}{75\%}\\&11615197&46649636&\\\hline
\end{tabular}
\end{table}

ESBMC with contractors had been evaluated on unreach-call property in $13$ subcategories in SV-COMP. 
Table~\ref{Table1} shows our comparison results with other BMC state-of-the-art tools. Our approach showed its effectiveness in multiple subcategories regarding score and resource consumption. In detail, we improved ESBMC incremental-BMC in \texttt{Arrays}, \texttt{ControlFlow}, \texttt{ECA}, \texttt{Heap}, and \texttt{Loops}. Moreover, Table~\ref{table3} shows how much memory contractors saved during the execution of these benchmarks. Although some categories consumed more memory with contractors, the total shows a combined decrease in memory usage that amount to $75$\%. 

\begin{tcolorbox}[floatplacement=ht,float]
The experimental results show that our method helped BMC efficiently reduce resource consumption, which includes a reduction of 75\% of memory usage while verifying $1$\% more benchmarks than our baseline, which thus successfully answers \textbf{EG1}.
\end{tcolorbox}

Our approach had shown its capabilities in these benchmarks by scoring the same as ESBMC incremental-BMC or even higher. We have the lead in subcategories \texttt{XCSP} and \texttt{AWS-C-Common}, and we are a close second in \texttt{Floats}. Also, our approach ranked third in our comparison of \texttt{Loops}, \texttt{Arrays}, and \texttt{Heap} categories. 
%AWS-C-Common subcategory represents real-world programs. We got promising results since we scored the highest in our comparison. 
This shows how valuable contractors are in real-world programs because we scored higher than other tools and decreased the verification time and memory. However, contractors could not improve the score in other categories, for it was not applicable in the current implementation. The cases where contractors could not be applied are due to our naive approach to detecting monotonicity and due to it only supporting a subset of operators (cf. Subsection~\ref{lim}).
%In contrast, we kept the same soundness and completeness ESBMC incremental-BMC has.

\begin{tcolorbox}[floatplacement=ht,float]
With the current scores, we can confirm that we have preserved the soundness (cf. Definition~\ref{def:soundness}) and completeness (cf. Definition~\ref{def:completeness}) of the BMC technique. The results show that the number of correct and incorrect results were maintained by a margin of 0.01\%, thus achieving our \textbf{EG2}.

%However, our approach suffered in \texttt{bitvectors} category mainly because of an overflow that our implementation cannot handle yet. In contrast, we scored the same results or higher in ESBMC incremental-BMC, thus achieving our \textbf{EG2}.
\end{tcolorbox}

One of the main issues in BMC is dealing with lengthy loops. Our approach focuses on dealing with loops and pruning the unnecessary steps to conclude the verification. \texttt{Loops} category has $14$ subcategories categorized the same as SV-COMP 2022 as shown in Table~\ref{Table2}. In particular, in four of these subcategories, our approach performed better than plain ESBMC. For example, we looked at the subcategory of \texttt{loops}; we got remarkable results compared to plain ESBMC. The same applies to \texttt{loop-new}, \texttt{loop-zilu}, \texttt{loops-crafted-1} and \texttt{nla-digbench-scaling}. We also got better scores and time consumption results than the champion tool Veriabs. In particular, in the subcategory \texttt{loop-simple}, \texttt{verifythis} and \texttt{nla-digbench-scaling}.

\begin{tcolorbox}[floatplacement=ht,float]
In the \texttt{loops} category, we find our approach shows improvement of ESBMC performance by at least $2$\% and achieves  \textbf{EG3}.
\end{tcolorbox}

However, we scored lower in \texttt{loop-acceleration} because some benchmarks timed out, which cost us $3$ scores. We also have some shortcomings in our approach: we do not deal with multiple lengthy loops in this version. That is why we got the same score in some subcategories. In general, we achieved the most critical objectives by saving CPU time, decreasing memory consumption and improving the scores. Additionally, we scored reasonably well in most categories, with $2$ being the lead. However, this version of our approach does not apply to some benchmarks, so we see some differences in scores. Also, we noticed our approach did not perform well in one particular benchmark and reported a false negative. The benchmark is in \texttt{bitvectors}, which represents the overflow problem. Another issue with the implementation is that we do not apply the contractors when it comes to an unsupported operator. This is also a reason for the score not improving in some subcategories. Moreover, the lack of accuracy in the interval analysis option in ESBMC hindered the performance of the contractors. Because the intervals were not accurate and the result of the contractor was not helpful to reduce the search space. In the future, we will work to overcome this limitation.

%----------------------------------------------------------------
\subsection{Threats to the validity}
\label{sec:TTTV}
%----------------------------------------------------------------

As mentioned in Section~\ref{sec:ProposedMethod}, we only apply the contractor if the contracted area is at the end of the loop. For example, given a variable $x=X_0$ that is incremented inside a non-deterministic loop $x++$ up to a max value of $X_{max}$, if the contractor result (e.g., $X = [X_0, X_{max}]$) was reduced from the upper bound (i.e., $X < X_n < X_{max}$, where $X < X_n$ is the contracted region) we can safely skip the instructions (iterations) that exceed the upper bound. However, suppose the contracted area was at the beginning of the loop (i.e., the reduced interval from the lower bound). In that case, we cannot guarantee the correctness of the verification results, as the method cannot reason about the minimum quantity of loop iterations. In the previous example, reducing from the lower bound (i.e. $X > X_n \wedge X < X_{max}$) cannot be contracted as it would skip instructions.

%For example, our program with one variable $x$ incremented in a loop may not lose any information; however, it may be impossible in other programs with multiple variables.
%Because skipping to that point in the program will result in some information loss. 
\begin{figure}
        \centering
        \includegraphics[width=0.33\columnwidth]{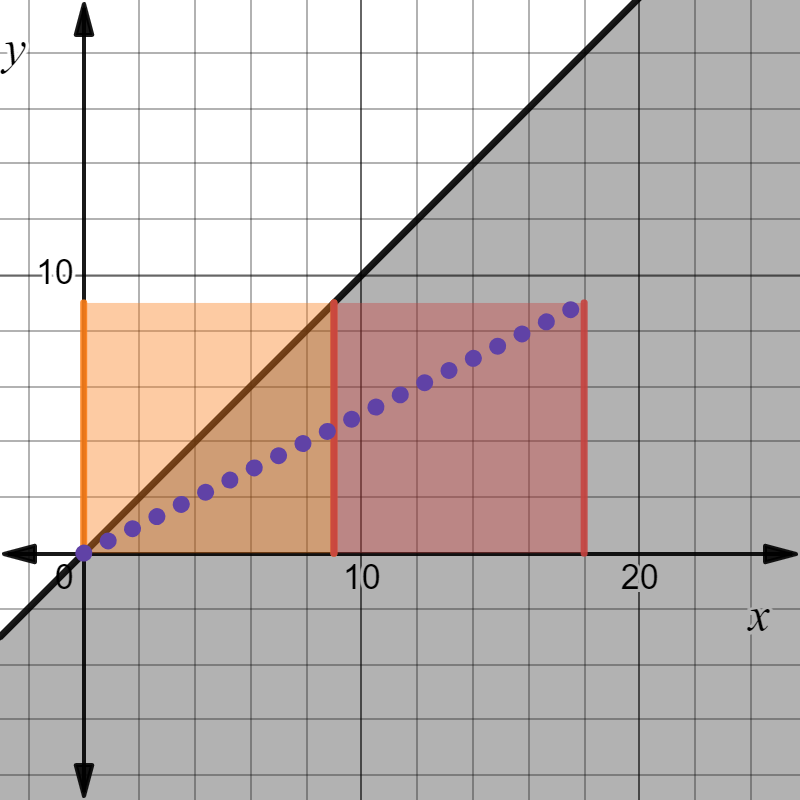}
        \includegraphics[width=0.33\columnwidth]{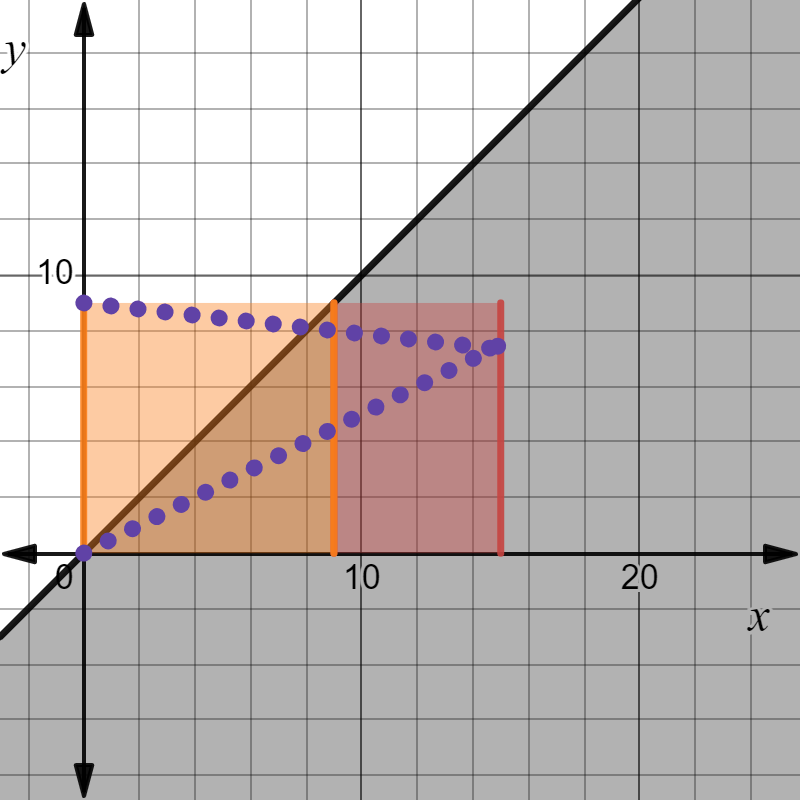}
        \caption{Monotone vs non-monotone.}
        \label{fig:mono}
    \end{figure}
    
Similarly, the program's monotonicity for variables also affects the verification results. For example, given a variable $y$ that may be non-deterministically incremented or decremented cannot be contracted. Illustrated in Figure~\ref{fig:mono} are two programs; one is monotone (here, $y$ would always increment), and the other is not (here, $y$ can increment and decrement at any state). Dots represent states, and blank lines and grey areas represent a property. Assume that we start from the origin point; the first program is safe to use an \textit{assume} directive to skip the states guaranteed to hold the property. However, that is not the case with the second program, and we will end up in the same situation as mentioned before, where the area reduced is at the beginning of the loop. Nevertheless, this time it is in the middle.

%----------------------------------------------------------------
\section{Related Work}
%----------------------------------------------------------------

One of the most known software verification frameworks for static analysis and verification for C/C++, and Java is CBMC~\cite{ckl2004,kroening2014cbmc}. CBMC is a bounded model checker that has been developed and maintained for over $18$ years. It implements bit-precise BMC to verify C programs. CBMC is part of CProver, which has the component \textit{Goto-Analyzer}~\cite{Yamaguchi2019ApplicationSystem}. It is an abstract interpreter and takes the GOTO-program as an input. It supports three types of abstract domain: \textit{constant}, \textit{variable sensitivity}, and \textit{dependency domain}. \textit{Goto-Analyzer} strength lies where it can analyze the interval, where it is guaranteed to hold and produce results as a success. Another tool for static analysis is Frama-C~\cite{cuoq2012frama,beyer2022static}. It is a framework that utilizes several plugins for static analysis of C code. One of these plugins is Evolved Value Analysis (EVA)~\cite{Buhler2020EvaPlugin}, which does the breakdown of each possible variable value and thus gives the interval(s) for the variables. 
It can analyze intervals accurately and produce a variable's calculated minimum and maximum value. Another tool with successful applications in the industry is Astree~\cite{Cousot2005Astree}. Astree is an abstract interpretation-based static program analyzer. It aims to prove the absence of run-time errors in C programs automatically. It uses an accurate over-approximation of intervals for variables, whether it is a floating-point or an abstract domain. It accounts for rounding errors in both domains. The use of interval methods is what distinguishes our work from Astree. 

There is also PIPS \cite{irigoin1991semantical} that was first introduced in $1988$. PIPS automatically detects medium- to large-grain parallelism in programs using techniques based on convex array regions. The main drawback of PIPS is that it only supports FORTRAN and C~\cite{Irigoin2014ret}. Like PIPS, PAGAI \cite{henry2012pagai} is also an automatic static analysis tool that uses the program's polyhedral abstraction. Both tools have been used to infer program invariants~\cite{rocha2017a}. APRON is another example of a library employed for static analysis of the program numerical variables using Abstract Interpretation~\cite{JeannetM09}. Such analysis aims to infer (inductive) invariants about those variables, such as $-2<=a-b<=c$, which should hold during any program execution.

Veriabs~\cite{darke2021veriabs, 8952452} is a tool that utilizes BMC, abstraction, and fuzzing to verify programs. Also, it employs a strategy selector that utilizes twelve different verification techniques. The selector will choose a technique based on code-structural and variable-data properties. Veriabs is the champion of SV-COMP 22 in Reach safety category. The runner-up in SV-COMP 22 is CPA Checker~\cite{beyer2011cpachecker} which is a tool that has configurable program analysis (CPA) to continuously improve with new verification components. CPA Checker is mainly built for researchers to conveniently implement new ideas and produce experimental results. 

What differentiates our approach from existing methods is the use of interval methods described in the interval analysis and methods field. All the studies motioned in our related work utilize the interval analysis techniques as described in static analysis. While Interval Analysis and Methods described in Section~\ref{sec:Background} share the same formal definitions and arithmetic, they do not share the same methods, in particular \textit{contractors}.

%---------------------------------------------------------------- 
\section{Conclusions and future work}
\label{conclusion}
%----------------------------------------------------------------

In Interval Analysis and Methods, contractors made using Interval Methods more practical in their applications by reducing the complexity of such methods. We advocate that contractors should have a similar effect for interval analysis applied to static analysis of programs. In this work, we extended the application of contractors to prune the search space of BMC techniques; the experimental results show its impact on open-source C benchmarks where the score was increased in some categories and reduced consumption of resources. In particular, interval methods via contractors made the incremental BMC verification faster and reduced memory usage by various orders of magnitude. Once we compare verification with and without the contractor, we can see a difference in the score ($32$ more benchmarks successfully verified) and resource consumption ($75$\% less memory consumed), especially when we prune the search space of programs that contain loops. Furthermore, compared to state-of-the-art tools in BMC, it also performed well, where we achieved the lead in $2$ subcategories. It preserved the soundness and completeness of the BMC technique by producing the same verification outcome. However, the approach needs more development to cover more cases and improve scores further. Implementing support for more operators and better interval analysis is top on the list of improvements that would have the highest impact on verification results in terms of score and resource consumption.

\bibliography{references.bib}
\bibliographystyle{ieeetr}
\clearpage

\end{document}